\documentclass[9pt, table, x11names, twocolumn]{article} 

\usepackage{preprint} 
\usepackage{svg}

\begin{document}

\togglecolumns 

\title{\textbf{Digital identity architectures: comparing goals and vulnerabilities}}

\date{\vspace{-5ex}}
\author[1]{Callum Mole}
\author[1]{Ed Chalstrey}
\author[1]{Peter Foster}
\author[1]{Tim Hobson}
\affil[1]{The Alan Turing Institute}

\twocolumn[ 
\begin{@twocolumnfalse} 

\maketitle
\begin{abstract}
   Digital identity systems have the promise of efficiently facilitating access to services for a nation's citizens while increasing security and convenience. There are many possible system architectures, each with strengths and weaknesses that should be carefully considered. This report first establishes a set of goals and vulnerabilities faced by any identity system, then evaluates the trade-offs of common digital identity architectures, principally comparing centralised and decentralised systems.
\end{abstract}
\vspace{10mm} 

\end{@twocolumnfalse}
] 

\thispagestyle{fancy} 

\section{Introduction}

\addreportpara

Across the world, access to basic services usually requires proof of identity. Healthcare, education, financial services, to name a few, all depend on verifiable attributes, yet many nations have weak identity systems \citep{worldprinciples}. A flawed identity system has stark consequences not only for the individual, who may be excluded from many services, but also for the nation state as a whole, that is unable to effectively develop or organise \citep{worldprinciples}. In the world today there exists an identity gap between countries that have well-formed identity systems developed over centuries, and countries that lack basic identity system infrastructure \citep[for example, a civil registry;][]{worldprinciples}. 

Digital identity systems, particularly those based on biometric data, are making significant headway in closing the identity gap. Leveraging digital innovations (e.g. biometrics; smartcards) could increase the security, accessibility, and convenience of identity systems \citep{worldprinciples}. There are, however, multiple ways to design digital identity systems, with no current best practice consensus. The critical design decisions that influence how the system operates (e.g. how users are authenticated) concern where identity information is held and how information is governed. At two extremes the data is held in a central location under the jurisdiction of a trusted entity (centralised), or held and managed by each individual user \citep[decentralised self-sovereign identity;][]{allen2016, kubach2020self}. Between these two extremes lie many possible architectures, each with its own trade-offs \citep{aiemworawutikul2019vulnerability}. This report evaluates these trade-offs, with respect to some fundamental goals that any digital identity systems should achieve. 

\section{Goals}

Before describing identity system architectures we first establish the criteria which we will use to evaluate a digital identity system. Though there is not currently a sufficiently general or widely accepted evaluation framework \citep{butlernational}, Cameron's \textit{Laws of Identity} \citep{cameron2005laws} is a popular choice when reviewing real-world systems \citep{panait2020identity,dunphy2018first,liu2020blockchain, chadwick2009federated}. Motivated by the lack of security and privacy in internet connections, Cameron's seven laws of identity relate principally to the privacy, usability, and security aspects of digital identity \citep{cameron2005laws}. 

Cameron's laws of identity do not consider infrastructure, limiting their applicability to contexts where the need for a sustainable identity system may override more abstract ideals. The World Bank Group's ID4D initiative has published ten principles for identification with a focus on sustainable development \citep{worldprinciples}. These principles, and the associated diagnostic framework \citep{worlddiagnostic}, extend many of the themes of Cameron's laws of identity to encompass practical considerations along the axes of inclusion, design, and governance. 

Here we build upon the work of Cameron and ID4D to outline clear common goals for identity systems to effectively function, with the aim of better evaluating the trade-offs faced by different identity architectures in the real world. The goals can be broadly grouped into three categories: \textbf{functional}, \textbf{privacy-related}, and \textbf{operational}. These categories will be discussed in turn. For a summary list of key goals mentioned see \textbf{Table~\ref{tab_goals}}.

\subsection{Functional Goals}

\textbf{Functional goals} are a set of minimum criteria that \textit{must} be met in order for a system to be an \textit{identity system}. Functionally a digital identity system is required to support three activities: \textbf{identification}, \textbf{authentication}, and \textbf{authorisation} \citep[][Figure 3, p 6]{goodell2019decentralised}. A successful identity system allows individuals to establish a \textit{credential} holding one or more specific identity attribute (e.g. residency status). In order to provide \textit{identification}, the credential must be verifiable at a later date (\textit{authentication}), so it can be used to \textit{authorise} access to services associated with the attributes it contains (e.g. voting). An example credential is the user's bank card number, which entitles the user to make a payment from their bank account. In this case, the credential issuer is the user's bank. When making a payment, authentication relies on a card payment organisation verifying the credential against the identity system. With the credential verified successfully, the user is entitled to use the credential to make a payment.

The functional goals require an ID system to have certain characteristics \citep{butlernational}. Authentication is only possible if the verifier is certain that any credentials establishing a user's identity are exclusive to the user, and have not been altered without due authorisation, or revoked. Therefore, the user needs to be \textit{uniquely bound} to their identity, and the data underpinning the identity needs to be \textit{tamper proof}, \textit{revocable} (e.g. in case of loss), and \textit{updateable} \citep[e.g. in case of change of user circumstance;][]{butlernational}. These properties are closely related to the elements of the classic information security triad: confidentiality, integrity, and availability \citep[][but see \citealt{lundgren2019defining} for critique]{bishop2005introduction}.  In an ID system context, authentication cannot occur if the appropriate resources and data are not \textit{available}, it needs to have \textit{integrity} so that one can be sure that credentials cannot be changed by malicious parties, and there needs to be \textit{confidentiality} so that data cannot be accessed and used by others \citep[][e.g. identity theft]{cameron2005laws}.

\subsection{Privacy Goals}

It is clear that privacy is essential for ID systems to effectively function, but \textit{how much} privacy there should be in an ID system is a hot societal issue. Indeed, the UN declaration of human rights Article 12 states that "No one shall be subjected to arbitrary interference with his privacy...Everyone has the right to the protection of the law against such interference or attacks" \citep{UNdec}, and privacy concerns have caused the legislative defeat of a number of proposed digital ID systems across the world \citep{mozillabringing}. Privacy can be considered critical to the able functioning of the system \citep{cameron2005laws}; data leaks pose a real threat to a system's ability to authenticate and authorise since unauthorised persons may now possess credentials. Some authors take a hard line on privacy, contending that the system should be designed in such a way that a user's sensitive data is never collected by other components of the identity system \citep{goodell2019decentralised}. However, the level of acceptable privacy is in some sense cultural, depending on a society's attitude towards trusted authorities and also the available infrastructure \citep[for example, countries with less established civil registries need to be more tightly bound to their identities upon use, potentially compromising privacy;][]{butlernational}, and there already exist ID systems where sensitive data is accessed, or traceable, by a central authority, for example, Nigeria's NIN \citep{NIN} and India's Aadhaar systems \citep{aadhaar}. Given the importance of privacy in the modern world one of the key questions around any ID system is not what functions an ID system should do, but how private it should be. 

In the ENISA position paper, privacy risk is defined to be `the unwanted disclosure of personal information and its subsequent misuse' \cite{naumann2008privacy}. These risks are considered to be fundamentally important as they represent a threat to the fundamental human rights of the citizen as enshrined in Article 8 of the European Convention on Human Rights. According to the ISO/IEC 15408 standard, the privacy protection goals include \textbf{anonymity}, \textbf{pseudonymity}, \textbf{unlinkability}, and \textbf{unobservability}. However, the privacy protection goals are not so well established compared to the classic triad of security protection goals. To help evaluate the privacy of an identity system, we refer to the privacy goals proposed by \cite{zwingelberg2011privacy}: \textbf{unlinkability}, \textbf{transparency}, and \textbf{intervenability}. Privacy means that users cannot have their actions traced (unlinkability), their personal data accessed in an unauthorised manner \citep[transparency; see also][]{cameron2005laws}, and there should also be the ability to interfere with any process in order to stop undesired actions from occurring (intervenability).

A related issue to unauthorised access of data is the principle of \textbf{minimal disclosure} \citep{cameron2005laws}. Privacy can be compromised by sharing data that is unnecessary for the use case, potentially compromising unlinkability.  For example, when purchasing alcohol in the UK, the only required data is whether one is over 18, not one's age or name. Similarly, the Global Privacy Standard (GPS) consists of ten privacy principles which considered to be the foundation of the user’s rights over the collection and use of their personal information \cite{cavoukian2010privacy}. 

\subsection{Operational Goals}

In addition to functional and privacy goals, there are practical considerations that any real-world system needs to contend with. Notwithstanding any theoretical justification of how a system should function, practically an ID system needs to be accessible by users over time \citep{worldprinciples}. We call goals relating specifically to the system's operation \textbf{Operational Goals}. A critical operational goal is \textit{sustainability} \citep{worldprinciples}. A system will fail if the demand of a system exceeds the available infrastructure and resources (e.g. it is not sufficiently \textit{cost effective} or \textit{efficient}). A system should also be future-proofed by allowing \textit{innovation} on dependent services \citep{cameron2005laws, worldprinciples}.

Perhaps the most fundamental operational goals is that the system needs to be usable by the population it serves \citep{ford2020identity, worldprinciples}. Usability can be considered from an infrastructural perspective, for example an Internet connection may be needed for authentication (system inaccessibility also compromises the functional goal of \textit{availability}). Usability should also be considered at the level of individual user acceptance. A system should be simple, understandable, and efficient to use \citep{cameron2005laws, worldprinciples, dhamija2008seven}, and trust should be built through due consideration of legal frameworks and oversight \citep{worldprinciples, mozillabringing}. One of the examples is the Pan-Canadian Trust Framework which contains principles of the desired identity ecosystem \cite{diacc2016trustcanada}. It includes principles such as transparency in governance and operation, user's choice and control, interoperability with international standards.

Debates on digital identity often focus around functional design and privacy \citep{microsoftdecentralized, goodell2019decentralised, kubach2020self, panait2020identity}, but the importance of operational considerations should not be underestimated. In the real-world operational goals are arguably a stronger determinant of how a system is designed or implemented than any of the aforementioned functional and privacy goals. Operational issues can arise due to (sometimes complex and unforeseeable) interactions between system designs and the context in which they are implemented. Furthermore, overlooking operational goals could lead to catastrophic failures, for example in both Peru \citep{peru2017reuben} and India \citep{india2018} administrative errors (e.g. appropriate legislation not being communication to front-line practitioners) resulted in blocking of essential services, and subsequent individual harm (even death).

Together, the functional, privacy, and operational goals constitute a framework with which to evaluate the effectiveness and resilience of digital identity systems.

\begin{table*}[htb]
\centering
\caption{\textbf{Summary description of key Digital Identity System Goals.}}
\begin{tabular}{p{3cm}p{3cm}p{7cm}} 
Goals & Category & Description \\
\midrule
Identification & Functional & user should be uniquely bound to their credentials \\ \greyrule
Authentication & Functional & credentials can be validated \\ \greyrule
Auditable & Functional & access and changes to data can be attributed \\ \greyrule
Non-repudiation & Functional & access and changes to data are undeniable \\ \greyrule
Authorisation & Functional & validation leads to use of entitled services \\ \greyrule
Availability & Functional / Operational & data should be available at all times \\ \greyrule
Integrity & Functional & data should be tamper proof and accurate \\ \greyrule
Modifiable & Functional & data needs to be revocable and updateable \\ \greyrule
Confidentiality & Functional / Privacy & data should not be accessible by unauthorised parties \\ \greyrule
Unlinkability & Privacy & users should not be traceable through their use of an ID system \\ \greyrule
Transparency & Privacy & users should be able to monitor how their data is being used \\ \greyrule
Intervenability & Privacy & undesirable processes should be stoppable \\ \greyrule
Minimal disclosure & Privacy & only the minimal relevant information for the transaction should be released to a service provider \\ \greyrule
Efficient & Operational & authentication and authorisation should be executed with minimum delay \\ \greyrule
Reliable & Operational & under the same conditions, the same outcomes should be achieved \\ \greyrule
Cost effective & Operational & running costs of an identity system should not exceed resources \\ \greyrule
Allows Innovation & Operational & an ID system needs to be able to flexibly adapt to future unforeseen changes in context \\ \greyrule
Comprehensible & Operational & a system should be simple to understand and use \\
\bottomrule
\end{tabular}
\label{tab_goals}
\end{table*}

\section{Vulnerabilities}

Having established a set of ID system goals we now consider the vulnerabilities posed to ID systems. Vulnerabilities are considered in a broad sense: things that can go wrong to prevent the system from achieving the goals set out above. These include \textbf{threats} arising from deliberately malicious behaviour, as well as accidental failure modes resulting from imperfect system design on a technical or operational level (e.g. producing misaligned incentives), or flawed implementation.

Each threat is assigned to one of three categories according to the level where the impact lies: individual, technological or organisational. The individual level refers to threats that can occur due to activities of individual persons within an otherwise well-functioning system. The technological level refers to vulnerabilities emerging from poorly designed or implemented systems. The organisational level refers to vulnerabilities that can occur when large organisations are responsible for running and maintaining the digital ID system. The distinction is intended as a conceptual aid, not an exact classification. The vulnerabilities from different levels interact (e.g. individual level vulnerabilities can be exacerbated by organisational level vulnerabilities), and a vulnerability categorisation does not also imply how the vulnerability could be mitigated (e.g. individual and organisation vulnerabilities could be mitigated by technical means). In the following sections the vulnerabilities associated with each level are described with reference to real-world examples. Table~\ref{tab_risks} collects these examples (and adds others), noting the vulnerabilities shown and the goals compromised.

\subsection{Individual-level vulnerabilities}

We consider two key ways that individual human activity can compromise a digital identity system: fooling the system (fraud), and user interaction failures.

Fraud is the use of an identity that does not legitimately belong to the user. Fraud appears in two forms \citep{FATF}: \textbf{impersonation} (or \textbf{identity theft}), where one person's identity is used by another, and \textbf{synthetic identity}, where fake or falsified information is used to create an identity that does not correspond to any existing person.

Fraud can occur in multiple ways. Individuals may be \textbf{fooled} or \textbf{coerced} into registering, de-registering or transacting within the system against their wishes, or indeed the system may be vulnerable to hacks that are enabled by corruption (e.g. developers sharing system vulnerabilities). Where deception is involved, it may be accomplished by running \textbf{spoof services} or \textbf{fake user interfaces} into which personal information (e.g. biometrics) is submitted. Fraudulently obtained encryption certificates may be used to convince the victim of the authenticity of the interface \citep{vanderbergrogue}.

An example of synthetic identity fraud comes from India's Aadhaar \citep{aadhaar}. In 2018, an Aadhaar hack sidestepped biometric enrolment procedures to allow unauthorised Aadhaar numbers to be generated. Installing the software patch was reportedly as simple as sourcing the code cheaply through WhatsApp groups provided by poorly paid enrolment operators \citep{aadhaarhack2018}.

Also at the individual level are failures in how the user interacts with the system. These failures result in individual exclusion, either from the system itself or from dependent services. Exclusion could occur due to \textbf{loss of credentials} such as a smart ID card, by the \textbf{lack of access} to the necessary digital technology or infrastructure (e.g. an Internet connection), or even by malicious action such as a deliberate denial of authenticity. Inability to authenticate due to loss of credentials or accessibility can be exacerbated by inappropriate alternative verification methods (see \textit{Technological Vulnerabilities}).

Exclusion could also occur when a system is \textbf{incomprehensible} to users, due to excessive complexity or reliance on esoteric technology. This may lead to outright exclusion but may also result in technical safeguards, such as those designed to protect user privacy, being neglected or used ineffectively \citep{dhamija2008seven, waldman2020cognitive}.

\subsection{Technological Vulnerabilities}

Flawed design and/or implementation at the technological level is an obvious source of vulnerability. Here we consider technological flaws leading to data being accessed by unauthorised persons, and more broadly technological failures that stop the system functioning.

Whenever an identity system, or indeed any software, requires the collection of personal information there is the threat of a \textbf{data breach}, that is the unauthorised extraction of data either by "insiders" (e.g. system administrators) or "outsiders" (i.e. hacking). Unauthorised access may be the result of a failure of encryption or key management, for example. In these cases, data may have been extracted at rest (from a database) or in transit, and the information in question may include personal data and/or system metadata (in the case of ID systems, this could be demographic or biometric data). Absence of effective access controls can also allow illegitimate \textbf{insertion or modification} of data within the system resulting in a loss of data integrity.

One example of a major data breach in recent history was that of Equifax, a US credit reporting agency, which in 2017 lost hundreds of millions of people's personally identifying data to hackers. The breach was enabled by a failure to apply a patch to a known vulnerability in a third party software \citep{equifax_2017}. 

A more subtle design vulnerability that compromises data security is \textbf{involuntary disclosure}, where the system provides unnecessary information in response to a particular request \citep[cf. zero-knowledge proofs where verification occurs without sharing of the verified data;][]{feige1988zero}. Even when individual user credentials successfully guarantee only selective disclosure, there is the vulnerability that information may be leaked if collections of credentials are created without ensuring a sufficient anonymity set (i.e. a large selection pool).

The most extreme consequence of inadequate technology may be complete \textbf{system failure}. There is risk of both intentional \textbf{denial of service}, such as a distributed denial of service (DDoS) attack across a wide range of servers \citep[e.g.][]{github2018}, and unintentional \textbf{loss of service} due to, for instance, a single point of failure losing power.

As a concrete example of denial of service, consider the cyber attacks suffered by Estonia in 2007. At the time much of Estonia's public digital infrastructure relied upon a single system remaining functional, so a coordinated attack via a large volume of automated online requests resulted in online banking and cash machines going offline and government bodies being unable to access email \citep{estonia2007}.

System failure may also occur indirectly, due to a failure of some \textbf{critical infrastructure} on which the system is dependent. Any system which is wholly dependent on communications or power networks will fail in their absence, but digital identity systems may also be dependent on additional network layers, such as a distributed ledger, whose long-term availability must be considered.

Another form of technological dependency that entails potential risk to the sustainability of the system is that of \textbf{vendor lock-in} \citep{cameron2005laws, worlddiagnostic}. Reliance on a single vendor of software or hardware (or on a small group) represents both a technological vulnerability, regarding long-term support for the platform, and an economic one. It is worth noting that vendor lock-in can lead to monopolisation of particular services. Monopoly entities can wield considerable power over a system, so vendor lock-in can also be considered an \textit{organisational vulnerability} (described in the next section), both as a control point of failure and as a powerful bad actor.

\subsection{Organisational Vulnerability}

The final category of vulnerabilities refers to the organisation(s) responsible for providing a digital identity system and therefore in control of its operation. Organisational vulnerabilities can emerge from asymmetric trust relationships, where users are forced to trust organisations with control of critical components of a digital ID system, such as holding sensitive data or authenticating users \citep{goodell2019decentralised}. In general, the organisational vulnerabilities described here take the form of data being used in ways that it was not originally intended for or that the user does not consent to. However, note that organisational control points also exacerbate some technological vulnerabilities such as inadequate security processes or power failures (see \textit{Technological vulnerabilities}). 

Whenever a particular organisation has influence over a digital ID system there is the concern of \textbf{function creep}, where the system is used for purposes beyond those originally intended. This is problematic because once the process of data collection is complete it is difficult to constrain the power of the data controller in deciding how to use it. This was the case in Pakistan where the national identity system has been linked to criminal databases to aid investigation \citep{pakistan2016}. 

A common and contentious example of function creep is the possibility of a central authority using the system as a means of user \textbf{surveillance}, either individually or \textit{en masse}. Such surveillance may take a variety of forms. One is the \textbf{surveillance of use}, via metadata such as authentication logs to determine who is using their ID for what purpose, and when and where they are doing so. Another form of surveillance involves the monitoring of register/deregister events to infer individual information via \textbf{set membership attacks}. A third form is the use of biometric data to track the physical movement of individuals, e.g. using automatic facial recognition. In some cases user surveillance is not the result of function creep \textit{per se}, but is a clear intention from the beginning, for example, in 2012 Mexico made biometrics mandatory citing bureaucratic inefficiency and personal safety \citep{velez2012}.

More generally, whenever personal information is collected there is a threat of \textbf{sharing of data} by the data controller without the users' consent. For example, Cambridge Analytica was able to obtain the personal data of millions of Facebook users \citep{confessore2018}. Cambridge Analytica was then able to use this data for \textbf{profiling}, a threat where subgroups are isolated by attributes of their identity for differential treatment. Profiling may be exacerbated by the \textbf{cross-linking of databases}, either by government or the private sector. This is easily facilitated in ID systems where a unique ID number and/or unitary biometric identity is used to access many services. Such systems are sometimes referred to as \textit{foundational identity} systems.

\begin{table*}[htb]
\centering
\caption{\textbf{Real world examples of ID system failure.}}
\begin{tabular}{p{3cm}p{3.5cm}p{4cm}p{4cm}} 
Real World Example & Source & Vulnerabilities Exposed & Goals compromised \\
\midrule
Aadhaar & \citealt{aadhaarhack2018} & Unauthorised insertion (of user enrolment data) & Integrity \\ \greyrule
Aadhaar  & \citealt{india2018} & Loss of service (accidental), exclusion, infrastructure failure & Availability, operability \\ \greyrule
Cambridge Analytica & \citealt{confessore2018} & Profiling, non-consensual data sharing & Unlinkability, Transparency, Intervenability \\ \greyrule
Capital One & \citealt{capitalone2019} & Data breach due to mis-configured AWS & Confidentiality, Intervenability \\ \greyrule
China & \citealt{china_2019} & Exclusion and Suppression of minorities through face recognition & Availability \\ \greyrule
e-Estonia & \citealt{estonia2007} & Denial of Service (intentional cyber attack) & Availability \\
\greyrule
e-Estonia  & \citealt{e-estoniaendgadget} & Impersonation (identity theft) & Confidentiality, Integrity \\ \greyrule Equifax & \citealt{equifax_2017} & Data breach & Confidentiality, Intervenability \\ \greyrule
Github & \citealt{github2018} & Denial of service (intentional DDoS attack) & Availability \\ \greyrule
Mexico & \citealt{velez2012} & Surveillance & Transparency, Intervenability, Unlinkability \\ \greyrule
Netherlands e-gov & \citealt{netherlands_diginotar} & Deceptive data collection (fraudalent encryption certificates) & Confidentiality \\ \greyrule
Pakistan & \citealt{pakistan2016} & Surveillance, Function Creep and non-consensual data sharing (id system linked to criminal database) & Transparency, Intervenability, Unlinkability \\ \greyrule
Peru & \ \citealt{peru2017reuben} & Loss of service (accidental), Exclusion & Availability \\ \greyrule
Refugees & \citealt{shoemaker2019identity} & Incomprehensibility of system for refugees & Transparency, Intervenability \\ \greyrule
Symantec & \citealt{symantec2017} & Deceptive data collection (faked certificates) & Confidentiality \\ 
\bottomrule
\end{tabular}
\label{tab_risks}
\end{table*}

\section{Digital Identity Architectures}

A key concept for describing a system's architecture is the extent to which it relies on a central controlling authority. A completely centralised system would hold all information in a single database and a single organisation would be responsible for governing how the information is used. Architectures diverge from this model geographically, by spreading data across multiple locations---a \textit{distributed} system---or organisationally, by spreading function and responsibility across either groups of service providers---a \textit{federated} system---or even to individual users---a \textit{decentralised} system. 

These concepts are useful for describing architectures but are not mutually exclusive, for example a system can be centralised in terms of governance but also distributed across many databases. The next section describes some example architectures. We then discuss architectural differences in terms of the goals and vulnerabilities described in previous sections, and also explore some of the opportunities for using privacy enhancing technologies (PETs) to help mitigate risks. Throughout, we focus on discussing the relative merits of more or less centralisation in the architecture design (summarised in \textbf{Figure~\ref{fig:centralised-vs-decentralised}}), but note that a single axis of centralisation cannot account for all the relevant differences between systems. 

We consider an identity system centralised if it is under the maintenance and control of a small number of influential organisations, referred to as platform operators. Platform operators are responsible for storing and handling the users' data. In the bank example given in the section on \textit{Functional Goals} the platform operators are the bank and card payment organisations. In centralised approaches participants depend on the platform operator's trustworthiness in order to conduct transactions successfully, securely, and privately. Often the trust relationship is mandatory, for example when making payments both users and service providers a forced to rely on particular card payment organisations (e.g. Visa or Mastercard).

A real-world example of a centralised digital ID system is the Aadhaar system in India \citep{aadhaar}. Note that the Aadhaar system is a legally accepted \textit{foundational} ID system, where one ID credential can be used to access most services. The Aadhaar system processes authentication requests in a government-run data centre, where credentials include biometrics in the form of iris and fingerprint data. Each time users access services the authentication process requires them to have their biometrics taken. Another example of a centralised system is NIN in Nigeria \citep{NIN} which is similarly based around state-controlled data and biometrics. However, even though both systems are strongly centralised it is worth noting that even within centralised systems there can be differences in how data is managed. For example, in Aadhaar the authentication data is replicated across two government data centres (so is in a sense \textit{distributed}),  whereas Nigeria's NIN uses one master data centre, from which subsets of data can be accessed by government organisations \citep[the service providers in this case;][]{butlernational}.

In contrast to centralised systems, which typically have only a single identity provider (and often a single identity), identity management can also be \textit{federated}, whereby subgroups of trusted entities recognise identities provided by one another \citep{maler2008venn, chadwick2009federated}. The critical difference is that federated systems, such an OpenID \citep{openid} and SAML \citep{saml}, support multiple identity providers, and users could distribute identity information across different domains.

Perhaps the most extreme form of a federated system is a truly decentralised system that altogether avoids the need for trusted third parties, with a user-centric approach \citep{allen2016, kubach2020self, lesavre2019taxonomic, dunphy2018decentralizing}. Decentralised systems where the user is the holder and manager of their own data are also called \textit{self-sovereign identity} systems \citep{allen2016, kubach2020self, lesavre2019taxonomic}. Decentralised systems tend also to be inherently distributed, often through the use of distributed ledger (a database synchronised across many locations) technology such as a blockchain \citep{liu2020blockchain, lesavre2019taxonomic, kubach2020self, dunphy2018first}. Note that while decentralised systems that do not use blockchain systems do exist (e.g. IOTA), in recent years there has been growing interest in blockchain-based solutions to achieving decentralised identity \citep{sovrinprotocol, liu2020blockchain, lesavre2019taxonomic, kubach2020self, dunphy2018decentralizing, dunphy2018first, ion,  goodell2019decentralised, volkov2020addressing} so throughout this report we focus on blockchain-based identity systems when discussing decentralisation. 

Decentralised systems are also often supported by standards for Decentralised Identifiers (DIDs), published by the Internet standards organisation World Wide Web Consortium (W3C), that consists of a URL that associates a person (or organisation or other entity) with a DID document \citep{w3cDID}. This enables the person who controls the DID to prove control of it without requiring permission from any central authority.

One example of a decentralised identity system that follows the W3C DID standard is Sovrin \citep{sovrinprotocol}. The Sovrin system avoids mandatory trust relationships with platform operators by providing a means for participants to maintain relationships directly with one another. Trusted peer-peer connections are achieved using DIDs and DID-mapped data, the latter which may include cryptographic verification keys and endpoint information. Both DIDs and DID-mapped data may be published to a distributed ledger. The ledger provides a historical record of data associated with a given DID, in a way which does not depend on centrally controlled data storage infrastructure. For the example of how this may work in practice consider an individual seeking communication with an insurance company. Based on knowledge of the insurance company's publicly published DID the individual requests the merchant's DID-mapped data from the ledger, allowing them to set up a secure communication channel with one another \citep{sovrinwhat}. For the specific purpose of maintaining their subsequent relationship both the individual and the insurance company may use newly generated DIDs to support asymmetrically encrypted communication. 

Besides using a distributed ledger to store DID-mapped data, Sovrin furthermore implements a decentralised approach by allowing participants to control where their credentials are stored \citep{sovrinwhat}. A user might install a software `wallet' application on their mobile device to store their credentials. Credentials are typically issued by one participant to another as cryptographically signed claims. For example, a user's name and address might be issued to the user as a claim by a local authority. When the user subsequently decides to enter a relationship with another participant, the user may decide to disclose any subset of claims as required by the other participant. An important feature of Sovrin is that by using zero knowledge proofs \citep{feige1988zero} and in combination with the distributed ledger, the user need not disclose the actual information contained in the claims. In this way, the user is able to prove that their home address has been verified, without disclosing their home address.

In summary, a key difference between centralised and decentralised systems is the extent to which hierarchical trust relationships are required. In centralised systems trusted third parties are essential control points of the system, and in return provide an efficient service. More decentralised systems attempt to avoid power imbalances but often have additional technological complexity. The next section discusses the strengths of each approach.

\section{Evaluation}

We now evaluate some of the digital identity architectures outlined in the previous section with respect to the goals and vulnerabilities that we identified earlier. In particular, we focus on the advantages and disadvantages of various digital identity system designs, noting the inevitable trade-offs that result from different levels of centralisation. We also note that centralisation can be measured along different axes. For example, an ID system which is ultimately controlled by a central authority may still decentralise certain aspects of its architecture, for instance by using a distributed database to mitigate certain technological vulnerabilities. 

In general, we see that more centralised systems benefit from increased efficiency and lower complexity in their implementation, allowing them to achieve \textit{Operational Goals}, but at the potential cost of amplifying certain \textit{Organisational Vulnerabilities}, with implications for users' security and privacy. On the other hand, more decentralised systems tend to fare well on \textit{Privacy Goals} by giving users ownership of (and responsibility for) their own data. This however brings with it an increased risk of user error leading to \textit{Individual-level Vulnerabilities}. These aspects, and others, are discussed in greater detail in the coming sections.
Finally, we briefly discuss potential benefits and challenges to the implementation of PETs in digital identity systems and suggest avenues for future work in this area.

\subsection{Goals}

This section compares how different identity system architectures meet the goals summarised in Table 1.

Digital ID systems that are currently deployed by nation states tend to have relatively centralised architectures. This is, however, something that can be thought of as on a spectrum, with examples such as the aforementioned Indian and Nigerian systems being at the centralised end and other nations such as Estonia \citep{eID} and Germany \citep{personalausweis} having a greater degree of decentralisation \citep{butlernational}. In practice, though highly decentralised systems that use blockchain technology, such as Sovrin (previously described), are the topic of much research \citep{panait2020identity, liu2020blockchain, lesavre2019taxonomic, dunphy2018first}, but yet to be deployed by any nation state. 

Some of the appeal of centralised systems are that they generally offer convenient and efficient services by leveraging existing societal trust relationships, skills, and potentially infrastructure, which pave the way for user acceptance and initial deployment. There is also the legislative appeal of having clearly assigned responsibility \citep{worldprinciples}, for example many would find it intuitive that a national ID system should be the responsibility of the state. Clearly assigned responsibility also may mean that the system is effectively resourced (for example providing salaried developers). In other words, centralised systems tend to do well on the the \textit{Operational Goals} of being efficient, cost effective, and accepted by users \citep{worlddiagnostic}.

A case can be made that, in developing countries where pre-existing civil registries of citizens are less common, the privacy trade-off associated with the use of biometrics in digital ID systems is balanced by the strong level of verification and authentication that is achieved by using this technology. In addition, easily modifiable central database(s) ease the administrative effort required to on-board people who are in urgent need of the services that a digital ID may entitle them to, which also have the benefit of easy revokability and updatability in cases of fraud or new relevant data on citizens respectively.

Finally, from the perspective of a citizen, one of the key potential benefits that centralised digital ID can provide is a clear authority to appeal to in situations where the use of a digital ID is disputed somehow \citep{worldprinciples}. As an example, imagine a scenario where you have been charged twice for a payment service provider. In this instance, the bank or card issuer can act as a trustworthy central authority to mediate the duplicate transaction dispute and potentially offer a refund.

\begin{figure*}
    \centering
    \includegraphics[width=0.7\textwidth]{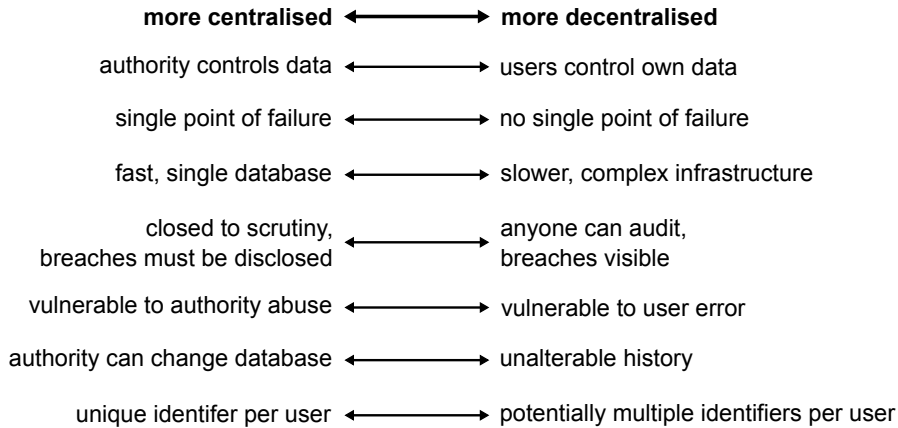}
    \caption{ Contrasting centralised and decentralised ID systems. }
    \label{fig:centralised-vs-decentralised}
\end{figure*}

At the other end of the spectrum each individual user is responsible for their own identity. Though the principles of decentralised systems are not new the number of  decentralised systems being developed is growing rapidly, in part due to the growing popularity of blockchain solutions to achieving decentralisation \citep{lesavre2019taxonomic, panait2020identity, kubach2020self, liu2020blockchain}. The obvious strengths of decentralised systems relate to the Privacy goals of \textit{transparency} and \textit{unlinkability}. Many proposed decentralised identity solutions, such as Sovrin \citep{sovrinprotocol} and The Decentralised Identity Foundation's ION \citep{ion, sidetree}, use a distributed ledger to keep an immutable record, meaning that data is easily auditable (and therefore more transparent). Further, decentralised systems can offer secure peer-peer connections \citep{lesavre2019taxonomic}, meaning that information exchange cannot be viewed by anyone not party to the secure channel (high \textit{confidentiality}). More user control over the providing information means that decentralised systems are more likely to meet the \textit{minimal disclosure} goal (though as we shall see later user control without understanding can also lead to vulnerabilities), whereby only the relevant information is provided to a service provider (zero-knowledge proof). Finally, decentralised systems tend to allow the proliferation of certificate and service providers, with an emphasis of achieving interoperability using open standards \citep[e.g. DIDs;][]{w3cDID}, which is a strong benefit for allowing future \textit{innovations} to be incorporated in the system.

\subsection{Vulnerabilities}

We now consider the extent to which each architectural approach addresses (or fails to address) the vulnerabilities (see Table~\ref{tab_risks}).

An astute reader may have noticed that the majority of the real-world examples in Table~\ref{tab_risks} are examples of centralised ID systems. This is due to the historical primacy of centralised systems. However, the availability of many examples also speaks to the vulnerabilities of highly centralised systems. 

Fundamentally, the risk of centralised systems lies in "the dubious assumption that some particular party or set of parties are universally considered trustworthy" \citep[][p 2]{goodell2019decentralised}. The existence of powerful platform operators means that many of the \textit{Organisational Risks} apply to any centralised system. Malicious powerful parties could repurpose data to their own aims (e.g. \textit{function creep}, \textit{non-consensual data sharing}), without the user being able to put the brakes on (\textit{intervenability}). For example, the central architecture of Aadhaar allowed a state-run dashboard whereby individuals were searchable by religion and caste due to combining sub-databases indexed with the Aadhaar number \citep[\textit{unlinkability};][]{mozillabringing}. Furthermore, if they choose to, the platform operators could obfuscate how the data is used, compromising \textit{transparency}. Crucially, the extent to which centralised systems meet the privacy goals is a policy choice on behalf of the platform operators rather than an inherent property of the system. 

Even if platform operators are well-intentioned their existence leads to identifiable control points. Control points exacerbate many technological risks. If a control point was to experience technological failure (such as power outage, or poorly setup access controls) the entire system is compromised. Notwithstanding the risk that central data stores usually have authorised personnel that could execute `insider' attacks from a place of privilege \citep[e.g.][]{aadhaarhack2018}, central data stores also invite `outside' attack since data breaches are potentially lucrative, for example profiteering from profiling. One could argue the intuition of keeping data stored in a central place that offers the best security around, but the existence of so many real-world examples suggests that this strategy is flawed and that data security is often compromised. 

It is precisely the risks caused by powerful platform operators that motivates the development of decentralised systems \citep{volkov2020addressing, goodell2019decentralised, lesavre2019taxonomic}. Therefore, the decentralised architecture mitigates many organisational and technological risks. Removing power from central authorities means that it is very difficult to repurpose data for one's own means (see \textit{Organisational Risks}) since in principle no particular entity has superior influence over the system. Furthermore, for blockchain solutions the data is transparent and auditable, or personal data remains securely private off-ledger \citep{sovrinwhat}. Without complete decentralisation it is possible to distribute power across federations, often with increased privacy due to pseudonymous use of services, yet each federation is nevertheless only as secure as its weakest member \citep{maler2008venn}.

Another benefit of more decentralised systems is that they are \textit{distributed}, often recruiting distributed ledgers that have many operating nodes, so are resistant to single (or even multiple) node failures. In contrast, while centralised systems can also be distributed (e.g. by replicating databases across many servers) this is a policy choice on behalf of the central authorities rather than an inherent property \citep{butlernational}. 

However, though decentralised systems attractively mitigate many risks they bring new risks to consider. Foremost of these risks is the additional technological complexities that gets shipped with decentralised systems. While more centralised systems (also federated systems) offer the convenience of a simple username and password, and often the ease of using the same credentials for many systems, some decentralised systems require considerable technical skill to run. The technological complexity is a recognised issue and some companies are working to develop user experiences that are as simple as entering a username and password into an app \citep{ion}, but there nevertheless remains a real risk of \textit{incomprehensibility} resulting in unwitting sharing of data or lack of user acceptance \citep{dunphy2018first, panait2020identity, dhamija2008seven, waldman2020cognitive}. 

A further user interaction error that comes with decentralised systems emerges from the user-centric approach. Though being responsible for one's own credentials is often considered advantageous there is currently no good solution for lost credentials \citep{volkov2020addressing}. A solution often proposed is to have third-party custodians \citep{volkov2020addressing, dunphy2018first}, but this introduces many of the organisational risks of relying on centralised authorities described above. Indeed, it is worth noting that it is very difficult in practice to be truly decentralised. Rather, truly decentralised systems currently only exist theoretically \citep[e.g.][]{goodell2019decentralised}, and most current practical decentralised identity management systems rely in some way on a trusted third party, either to provide resource (such as running the ledger) or trusted services \citep{goodell2019decentralised, dunphy2018decentralizing, dunphy2018first, panait2020identity, aiemworawutikul2019vulnerability}. 

In addition to the decentralisation problem of eliminating control points of trust, there are criticisms specific to blockchain-based solutions that the technical aspects of the network itself are not entirely decentralised. The compute required for proof of work, the consensus mechanism for adding blocks to the ledger, means that in practice the majority of proof of work is done by only a handful of nodes \citep{gencer2018decentralization}. Some authors have suggested this makes some blockchains vulnerable to 51\% attacks, where one or a pool of miners have the majority of mining power so can manipulate consensus, compromising immutability \citep{dunphy2018decentralizing, panait2020identity, blake2017}. One can avoid this problem by employing permissioned distributed ledgers, where only a trusted pool take part in the consensus \citep[e.g. The Sovrin Network;][]{sovrinprotocol}, but then one is once again at the mercy of central authorities.

Due to the compute required for consensus, some blockchain-based decentralised systems may operate more slowly than their centralised counterparts. For example, \cite{volkov2020addressing} reports that the delay between a transaction occurring and it being registered on the bitcoin blockchain is substantially slower than it would be for other kinds of database, the result being that bitcoin achieves a throughput rate of 7 transactions/sec, compared with Visa's 2,000. Furthermore, writing transactions to the blockchain is costly, and as the ledger grows so does the storage and compute requirements \citep{panait2020identity}. These properties cause scalability concerns (note that writing to the blockchain could be seen as a single point of failure) that do not come with centralised solutions. There is, however, proposed solutions for improving throughput, namely having a second layer that sits atop the blockchain layer and integrates operations in some way, resulting in less transactions to the blockchain \citep{lesavre2019taxonomic}.

\subsection{Opportunities and Challenges for PETs in digital identity systems}

Having provided an overview of the trade-offs of different architectures we now briefly discuss the potential benefits of PETs as identified in a recent Royal Society report \citep{royalprivacy}.  

In general, due to the sensitive nature of identity, \textit{encryption} of data at rest and in transit should be standard practice in all designs. Secure encryption provides some protection from data leaks. Furthermore, if an identity service is outsourced, for example on an untrusted cloud provider's platform, the additional protection against insider attacks from the platform's administrators is possible using \textit{secure enclave technology}. However, this brings with it the additional burden of managing the keys for use of the enclave.

There are a number of operations and management activities associated with running a digital identity system that require logging, and analysis of logs can certainly benefit from the use of \textit{Differential Privacy} techniques. This apples across all architectural design choices. We note that Differential Privacy requires \textit{a priori} knowledge of the queries that will be made, so that the relevant rate limits and statistical protection (epsilon) can be computed. These limits also constrain how many different administrators (or roles) are allowed to make such queries, so role based access control has to be managed carefully. 

\textit{Secure multiparty computation} (MPC) can offer privacy preserving computation in a distributed setting. The obvious example is the application to queries such as "is this person's age greater than X", where the service may store the date of birth of the person, but we wish to keep the actual date of birth private and the result of the query encrypted (the binary result would be available only to the parties involved). There are many other possible uses of MPC in identity systems that could be explored in future. Each needs careful design, though MPC use for private set intersection seems another candidate.

\textit{Homomorphic encryption} is another PET that has obvious application here. For example, if we wish to store biometric data securely in a third party centralised service, and allow confidential verification of users' biometrics, without revealing the result, then a fully homomorphic encryption (FHE) function can be constructed to do this. The performance limitations of FHE are probably not prohibitive for this sort of use case. In principle, one could also implement the log processing mentioned above using FHE, although in practice this might be computationally very expensive. Perhaps a hybrid of differential privacy and somewhat homomorphic encryption could be devised in that case.

\section{Summary}

We have discussed different digital identity system architectures with respect to goals they should meet and risks they should avoid. The key differences can be usefully abstracted along the axis of more centralised at one end and more decentralised at the other (Figure~\ref{fig:centralised-vs-decentralised}). 

Highly centralised systems benefit from ease of governance and use, but any system that has influential entities (including federated systems) is exposed to risks associated with potential misuse of data or control point failures. In contrast, highly decentralised systems have stronger privacy credentials by design and are more resilient, yet face open issues about scalability, governance, and usability. 

Though theoretically a completely decentralised system attractively avoids hierarchical trust relationships and promotes resilience \citep{goodell2019decentralised}, the practical benefits of centralisation, especially in terms of authority, resource, and user acceptance, means that some element of centralisation is likely in implemented digital id systems. 

The pertinent question, then, is how can we combine the strengths from both architectures? It may be possible, for example, to limit the organisation risks of powerful entities with the appropriate regulations and legislation \citep{worlddiagnostic}, or increase unlinkability of centralised architectures through the use of privacy-enhancing technologies \citep{royalprivacy}. In our penultimate section we provide a brief overview of how PETs could benefit digital identity systems. Future work will consider in detail the potential for applying PETs to different architectures. 

\section{Acknowledgements}
The authors gratefully acknowledge helpful discussions with Jon Crowcroft, Lydia France, Sam Greenbury, Luke Hare, Carsten Maple, and Pamela Wochner.

This work was supported, in whole or in part, by the Bill \& Melinda Gates Foundation [INV-001309]. Under the grant conditions of the Foundation, a Creative Commons Attribution 4.0 Generic License has already been assigned to the Author Accepted Manuscript version that might arise from any submission.

\FloatBarrier
\def\UrlBreaks{\do\/\do-} 
\newpage
\bibliographystyle{apalike}
\bibliography{bibliography}

\begin{thebibliography}{}

\bibitem[Aiemworawutikul et~al., 2019]{aiemworawutikul2019vulnerability}
Aiemworawutikul, W., Datla, M.~V., Lee, J. C.~S., Wen, T., and Zhang, Y.
  (2019).
\newblock Vulnerability assessment in national identity services.

\bibitem[Allen, 2016]{allen2016}
Allen, C. (2016).
\newblock The path to self-sovereign identity.
\newblock
  \url{http://www.lifewithalacrity.com/2016/04/the-path-to-self-soverereign-identity.html}.

\bibitem[Azeem, 2016]{pakistan2016}
Azeem, M. (2016).
\newblock {Nadra-linked criminal database becomes operational}.
\newblock Dawn: \url{https://www.dawn.com/news/1273886}.

\bibitem[Bhardwaj, 2018]{india2018}
Bhardwaj, M. (2018).
\newblock {For India's poorest, an ID card can be the difference between life
  and death}.
\newblock Reuters:
  \url{https://fr.reuters.com/article/us-india-election-starvation-analysis-idUSKCN1LS0HS}.

\bibitem[Bishop, 2005]{bishop2005introduction}
Bishop, M. (2005).
\newblock Introduction to computer security.

\bibitem[{BMI}, 2020]{personalausweis}
{BMI} (2020).
\newblock Personalausweis.
\newblock \url{https://www.personalausweisportal.de}.

\bibitem[Butler et~al., 2020]{butlernational}
Butler, D., Dzhamtyrova, R., Hicks, C., and Pooranian, Z. (2020).
\newblock National digital identity architectures in context.
\newblock in preparation.

\bibitem[Byler, 2019]{china_2019}
Byler, D. (2019).
\newblock {China's hi-tech war on its Muslim minority}.
\newblock Guardian article:
  \url{https://www.theguardian.com/news/2019/apr/11/china-hi-tech-war-on-muslim-minority-xinjiang-uighurs-surveillance-face-recognition}.

\bibitem[Cameron, 2005]{cameron2005laws}
Cameron, K. (2005).
\newblock The laws of identity.
\newblock Microsoft Corp.
  \url{https://www.identityblog.com/stories/2005/05/13/TheLawsOfIdentity.pdf}.

\bibitem[Cavoukian, 2010]{cavoukian2010privacy}
Cavoukian, A. (2010).
\newblock Privacy by design: the 7 foundational priniciples.

\bibitem[Chadwick, 2009]{chadwick2009federated}
Chadwick, D.~W. (2009).
\newblock Federated identity management.
\newblock In {\em Foundations of security analysis and design V}, pages
  96--120. Springer.
\newblock \url{https://kar.kent.ac.uk/30609/1/FederatedIdManChapter.pdf}.

\bibitem[Charette, 2011]{netherlands_diginotar}
Charette, R. (2011).
\newblock {DigiNotar Certificate Authority Breach Crashes e-Government in the
  Netherlands}.
\newblock IEEE Spectrum blog:
  \url{https://spectrum.ieee.org/riskfactor/telecom/security/diginotar-certificate-authority-breach-crashes-egovernment-in-the-netherlands}.

\bibitem[Cluley, 2018]{github2018}
Cluley, G. (2018).
\newblock {GitHub was hit by the most powerful DDoS attack in history}.
\newblock blog:
  \url{https://grahamcluley.com/github-hit-powerful-ddos-attack-history/}.

\bibitem[Confessore, 2018]{confessore2018}
Confessore, N. (2018).
\newblock {Cambridge Analytica and Facebook: The Scandal and the Fallout so
  far}.
\newblock New York Times:
  \url{https://www.nytimes.com/2018/04/04/us/politics/cambridge-analytica-scandal-fallout.html}.

\bibitem[{Decentralized Identity Foundation}, 2020a]{ion}
{Decentralized Identity Foundation} (2020a).
\newblock {ION}.
\newblock \url{https://github.com/decentralized-identity/ion}.

\bibitem[{Decentralized Identity Foundation}, 2020b]{sidetree}
{Decentralized Identity Foundation} (2020b).
\newblock Sidetree.
\newblock \url{https://github.com/decentralized-identity/sidetree}.

\bibitem[Dhamija and Dusseault, 2008]{dhamija2008seven}
Dhamija, R. and Dusseault, L. (2008).
\newblock The seven flaws of identity management: Usability and security
  challenges.
\newblock {\em IEEE Security \& Privacy}, 6(2):24--29.

\bibitem[{DIACC Trust Framework Expert Committee}, 2016]{diacc2016trustcanada}
{DIACC Trust Framework Expert Committee} (2016).
\newblock {Pan-Canadian Trust Framework Overview}.
\newblock
  \url{https://diacc.ca/wp-content/uploads/2016/08/PCTF-Overview-FINAL.pdf}.

\bibitem[Dunphy et~al., 2018]{dunphy2018decentralizing}
Dunphy, P., Garratt, L., and Petitcolas, F. (2018).
\newblock Decentralizing digital identity: Open challenges for distributed
  ledgers.
\newblock In {\em 2018 IEEE European Symposium on Security and Privacy
  Workshops (EuroS\&PW)}, pages 75--78. IEEE.

\bibitem[Dunphy and Petitcolas, 2018]{dunphy2018first}
Dunphy, P. and Petitcolas, F.~A. (2018).
\newblock A first look at identity management schemes on the blockchain.
\newblock {\em IEEE Security \& Privacy}, 16(4):20--29.
\newblock \url{https://arxiv.org/pdf/1801.03294.pdf}.

\bibitem[{e-Estonia}, 2020]{eID}
{e-Estonia} (2020).
\newblock eid.
\newblock \url{https://www.id.ee/}.

\bibitem[FATF, 2020]{FATF}
FATF (2020).
\newblock Guidance on digital identity.
\newblock Technical report, FATF, Paris.
\newblock
  \url{www.fatf-gafi.org/publications/documents/digital-identity-guidance.html}.

\bibitem[Feige et~al., 1988]{feige1988zero}
Feige, U., Fiat, A., and Shamir, A. (1988).
\newblock Zero-knowledge proofs of identity.
\newblock {\em Journal of cryptology}, 1(2):77--94.

\bibitem[Ford, 2020]{ford2020identity}
Ford, B. (2020).
\newblock Identity and personhood in digital democracy: Evaluating inclusion,
  equality, security, and privacy in pseudonym parties and other proofs of
  personhood.
\newblock {\em arXiv preprint arXiv:2011.02412}.

\bibitem[Fruhlinger, 2020]{equifax_2017}
Fruhlinger, J. (2020).
\newblock {Equifax data breach FAQ: What happened, who was affected, what was
  the impact?}
\newblock CSO blog:
  \url{https://www.csoonline.com/article/3444488/equifax-data-breach-faq-what-happened-who-was-affected-what-was-the-impact.html}.

\bibitem[Gencer et~al., 2018]{gencer2018decentralization}
Gencer, A.~E., Basu, S., Eyal, I., Van~Renesse, R., and Sirer, E.~G. (2018).
\newblock Decentralization in bitcoin and ethereum networks.
\newblock In {\em International Conference on Financial Cryptography and Data
  Security}, pages 439--457. Springer.
\newblock \url{https://arxiv.org/abs/1801.03998}.

\bibitem[Goodell and Aste, 2019]{goodell2019decentralised}
Goodell, G. and Aste, T. (2019).
\newblock A decentralised digital identity architecture.
\newblock {\em Frontiers in Blockchain}, 10.
\newblock \url{doi: 10.3389/fbloc.2019.00017}.

\bibitem[Goodin, 2017]{symantec2017}
Goodin, D. (2017).
\newblock {Already on probation, symantec issues more illegit HTTPS
  certificates}.
\newblock Arstechnica blog:
  \url{https://arstechnica.com/information-technology/2017/01/already-on-probation-symantec-issues-more-illegit-https-certificates/}.

\bibitem[Hall, 2017]{blake2017}
Hall, B. (2017).
\newblock 5 identity problems blockchain doesn’t solve.
\newblock Medium:
  \url{https://medium.com/@blake_hall/5-identity-problems-blockchain-doesnt-solve-ed4badb94398}.

\bibitem[Kak et~al., 2020]{mozillabringing}
Kak, A., Ben-Avie, J., Munyua, A., and Tiwari, U. (2020).
\newblock Bringing openness to identity.
\newblock White paper, Mozilla.
\newblock
  \url{https://blog.mozilla.org/netpolicy/files/2020/01/Mozilla-Digital-ID-White-Paper.pdf}.

\bibitem[Khaira et~al., 2018]{aadhaarhack2018}
Khaira, R., Sethi, A., and Sathe, G. (2018).
\newblock {UIDAI's Aadhaar software hacked, ID database compromised, experts
  confirm}.
\newblock Huffington post blog:
  \url{https://www.huffingtonpost.in/2018/09/11/uidai-s-aadhaar-software-hacked-id-database-compromised-experts-confirm_a_23522472/}.

\bibitem[Krebs, 2019]{capitalone2019}
Krebs, B. (2019).
\newblock {Capital One Data Theft Impacts 106m people}.
\newblock KrebsonSecurity blog:
  \url{https://krebsonsecurity.com/tag/capital-one-breach/}.

\bibitem[Kubach et~al., 2020]{kubach2020self}
Kubach, M., Schunck, C.~H., Sellung, R., and Ro{\ss}nagel, H. (2020).
\newblock Self-sovereign and decentralized identity as the future of identity
  management?
\newblock {\em Open Identity Summit 2020}.
\newblock
  \url{https://dl.gi.de/bitstream/handle/20.500.12116/33180/proceedings-03.pdf}.

\bibitem[Lesavre et~al., 2020]{lesavre2019taxonomic}
Lesavre, L., Varin, P., Mell, P., Davidson, M., and Shook, J. (2020).
\newblock A taxonomic approach to understanding emerging blockchain identity
  management systems.
\newblock {\em arXiv preprint arXiv:1908.00929}.

\bibitem[Liu et~al., 2020]{liu2020blockchain}
Liu, Y., He, D., Obaidat, M.~S., Kumar, N., Khan, M.~K., Choo, K.-K.~R., et~al.
  (2020).
\newblock Blockchain-based identity management systems: A review.
\newblock {\em Journal of Network and Computer Applications}, page 102731.

\bibitem[Lundgren and M{\"o}ller, 2019]{lundgren2019defining}
Lundgren, B. and M{\"o}ller, N. (2019).
\newblock Defining information security.
\newblock {\em Science and engineering ethics}, 25(2):419--441.

\bibitem[Maler and Reed, 2008]{maler2008venn}
Maler, E. and Reed, D. (2008).
\newblock The venn of identity: Options and issues in federated identity
  management.
\newblock {\em IEEE Security \& Privacy}, 6(2):16--23.

\bibitem[McGuinness, 2017]{estonia2007}
McGuinness, D. (2017).
\newblock {How a cyber attack transformed Estonia}.
\newblock BBC news: \url{https://www.bbc.com/news/39655415}.

\bibitem[Microsoft, 2018]{microsoftdecentralized}
Microsoft (2018).
\newblock Decentralized identity.
\newblock White paper.
\newblock
  \url{https://query.prod.cms.rt.microsoft.com/cms/api/am/binary/RE2DjfY}.

\bibitem[Moon, 2017]{e-estoniaendgadget}
Moon, M. (2017).
\newblock Estonia freezes resident id cards due to security flaw.
\newblock Endgadget blog:
  \url{https://www.engadget.com/2017-11-04-estonia-freezes-resident-id-cards-security-flaw.html}.

\bibitem[{National Identity Management Commission}, 2020]{NIN}
{National Identity Management Commission} (2020).
\newblock {NIN homepage}.
\newblock \url{https://www.nimc.gov.ng/}.

\bibitem[Naumann and Hogben, 2008]{naumann2008privacy}
Naumann, I. and Hogben, G. (2008).
\newblock Privacy features of european eid card specifications.
\newblock {\em Network Security}.

\bibitem[Oasis, 2007]{saml}
Oasis (2007).
\newblock Security assertion markup language (saml).

\bibitem[{OpenID Foundation}, 2007]{openid}
{OpenID Foundation} (2007).
\newblock Openid authentication 2.0.

\bibitem[Panait et~al., 2020]{panait2020identity}
Panait, A.-E., Olimid, R.~F., and Stefanescu, A. (2020).
\newblock Identity management on blockchain--privacy and security aspects.
\newblock {\em arXiv preprint arXiv:2004.13107}.

\bibitem[Reuben and Carbonari, 2017]{peru2017reuben}
Reuben, W. and Carbonari, F. (2017).
\newblock Identification as a national priority: The unique case of peru.
\newblock Working paper 454, Center for Global Development.
\newblock
  \url{https://www.cgdev.org/publication/identification-national-priority-unique-case-peru}.

\bibitem[Shoemaker et~al., 2019]{shoemaker2019identity}
Shoemaker, E., Kristinsdottir, G.~S., Ahuja, T., Baslan, D., Pon, B., Currion,
  P., Gumisizira, P., and Dell, N. (2019).
\newblock Identity at the margins: examining refugee experiences with digital
  identity systems in lebanon, jordan, and uganda.
\newblock In {\em Proceedings of the 2nd ACM SIGCAS Conference on Computing and
  Sustainable Societies}, pages 206--217.
\newblock \url{http://www.nixdell.com/papers/2019-refugee-COMPASS.pdf}.

\bibitem[{Sovrin Foundation}, 2018]{sovrinprotocol}
{Sovrin Foundation} (2018).
\newblock Sovrin: A protocol and token for self-sovereign identity and
  decentralized trust.
\newblock White paper.
\newblock
  \url{https://sovrin.org/wp-content/uploads/2018/03/Sovrin-Protocol-and-Token-White-Paper.pdf}.

\bibitem[{The Royal Society}, 2019]{royalprivacy}
{The Royal Society} (2019).
\newblock Protecting privacy in practice: The current use, development and
  limits of privacy enhancing technologies in data analysis.
\newblock Technical report.
\newblock
  \url{https://royalsociety.org/topics-policy/projects/privacy-enhancing-technologies}.

\bibitem[Tobin, 2018]{sovrinwhat}
Tobin, A. (2018).
\newblock Sovrin: What goes on the ledger?
\newblock White paper, Evernym, Sovrin Foundation.
\newblock
  \url{https://www.evernym.com/wp-content/uploads/2017/07/What-Goes-On-The-Ledger.pdf}.

\bibitem[{UN General Assembly}, 1948]{UNdec}
{UN General Assembly} (1948).
\newblock Universal declaration of human rights.
\newblock \url{https://www.refworld.org/docid/3ae6b3712c.html}.

\bibitem[{Unique Identification Authority of India}, 2020]{aadhaar}
{Unique Identification Authority of India} (2020).
\newblock {UIDAI official homepage}.
\newblock \url{https://uidai.gov.in}.

\bibitem[Vanderburg, 2012]{vanderbergrogue}
Vanderburg, E. (2012).
\newblock Threat of rogue certificate authorities.
\newblock TDCI.
  \url{https://www.tcdi.com/the-threat-of-rogue-certificate-authorities/}.

\bibitem[Volkov, 2020]{volkov2020addressing}
Volkov, A. (2020).
\newblock Addressing the challenges facing decentralized identity systems.
\newblock {\em iSCHANNEL}, 15(1):10--15.

\bibitem[Vélez, 2012]{velez2012}
Vélez, A. (2012).
\newblock Insecure identities. the approval of a biometric id card in mexico.
\newblock {\em Surveillance and Society}, 10(1):42--50.
\newblock \url{https://doi.org/10.24908/ss.v10i1.4147}.

\bibitem[{W3C}, 2020]{w3cDID}
{W3C} (2020).
\newblock Decentralised identifiers (dids) v1.0.
\newblock \url{https://www.w3.org/TR/did-core/}.

\bibitem[Waldman, 2020]{waldman2020cognitive}
Waldman, A.~E. (2020).
\newblock Cognitive biases, dark patterns, and the ‘privacy paradox’.
\newblock {\em Current opinion in psychology}, 31:105--109.

\bibitem[{World Bank Group}, 2018a]{worlddiagnostic}
{World Bank Group} (2018a).
\newblock Guidelines for id4d diagnostics.
\newblock Working paper, Washington, D.C.
\newblock
  \url{http://documents.worldbank.org/curated/en/213581486378184357/Principles-on-identification-for-sustainable-development-toward-the-digital-age}.

\bibitem[{World Bank Group}, 2018b]{worldprinciples}
{World Bank Group} (2018b).
\newblock Principles on identification for sustainable development : Toward the
  digital age.
\newblock Working paper, Washington, D.C.
\newblock
  \url{http://documents1.worldbank.org/curated/en/370121518449921710/Guidelines-for-ID4D-Diagnostics.pdf}.

\bibitem[Zwingelberg and Hansen, 2011]{zwingelberg2011privacy}
Zwingelberg, H. and Hansen, M. (2011).
\newblock Privacy protection goals and their implications for eid systems.
\newblock In {\em IFIP PrimeLife International Summer School on Privacy and
  Identity Management for Life}, pages 245--260. Springer.

\end{thebibliography}


\end{document}